\documentclass[12pt]{iopart}
\newcommand {\rhovec}{\ensuremath \boldsymbol{\rho}}
\usepackage{graphicx,iopams}

\begin{document}
\title[Comment on `Simulating thick atmospheric turbulence in the lab...']{Comment on `Simulating thick atmospheric turbulence in the lab with application to orbital angular momentum communication'}
\author{Jeffrey H Shapiro}
\address{Research Laboratory of Electronics,
Massachusetts Institute of Technology, Cambridge, MA 02139, USA}
\ead{jhs@mit.edu}

\begin{abstract}
Recently, Rodenburg \em et al\/\rm\ (2014 New J. Phys.\@ {\bf 16} 033020) presented an approach for simulating propagation over a long path of uniformly distributed Kolmogorov-spectrum turbulence by means of a compact laboratory arrangement that used two carefully placed and controlled spatial light modulators.  We show that their simulation approach mimics the behavior of plane-wave propagation, rather than general beam-wave propagation.  Thus, the regime in which their orbital angular momentum (OAM) cross-talk results accurately represent the behavior to be expected in horizontal-path propagation through turbulence may be limited to collimated-beam OAM modes whose diameters are sufficient that turbulence-induced beam spread is negligible.
\end{abstract}

\maketitle 
\section{Introduction}
The use of orbital angular momentum (OAM) beams in free-space optical (FSO) communication has attracted considerable interest of late, both for increasing the data rate of classical communications \cite{Padgett,Ren1} and the secret-key rate of quantum communications \cite{Malik,Gauthier}.  Of course, such OAM FSO systems are subject to performance degradation arising from atmospheric turbulence, thus prompting work to understand the nature of that degradation \cite{Paterson,Boyd,Ren2}, and how adaptive optics might mitigate it \cite{Ren3}.  Almost exclusively, however, studies of  turbulence effects on OAM propagation, such as \cite{Paterson,Boyd,Ren2,Ren3}, have presumed thin, phase-screen turbulence in the vicinity of the receiver pupil.   Consequently, these works do not properly characterize the effects that would be encountered in propagation over a long path of uniformly-distributed turbulence, i.e., thick turbulence.  Recently, Rodenburg \em et al\/\rm\ \cite{Rodenburg} presented an approach for simulating propagation over long path of uniformly distributed Kolmogorov-spectrum turbulence by means of a compact laboratory arrangement that used two carefully placed and controlled spatial light modulators.  We show that their simulation approach mimics the behavior of plane-wave propagation, rather than general beam-wave propagation.  Thus, the regime in which their OAM cross-talk results accurately represent the behavior to be expected in horizontal-path propagation through turbulence may be limited to collimated-beam OAM modes whose diameters are sufficient that turbulence-induced beam spread is negligible.  

We begin, in section~2, by showing how the extended Huygens-Fresnel principle \cite{Shapiro} can be used to characterize the average cross-talk between OAM modes that have propagated through thick turbulence.  Then, in section~3, we demonstrate that \cite{Rodenburg} may only capture that behavior when turbulence-induced beam spread can be neglected for their collimated-beam OAM modes \cite{RodenburgPrivate}.  There, we also show that this no beam-spread condition is barely satisfied by the simulation parameters they establish for 785-nm-wavelength light to propagate over a 1-km-long path through uniformly distributed $C_n^2 = 1.8\times 10^{-14}\,{\rm m}^{-2/3}$ Kolmogorov-spectrum turbulence from an 18.2-cm-diameter transmit pupil to an 18.2-cm-diameter receive pupil.

\section{Cross-Talk Characterization via the Extended Huygens-Fresnel Principle}
Let $\{\Psi_\ell(\rhovec)\}$, for $\rhovec = (x,y)$, be the complex field envelopes for a set of wavelength-$\lambda$ orthonormal OAM modes on the circular transmitter pupil $\mathcal{A}_0 = \{\rhovec: |\rhovec|\le D/2\}$ in the $z=0$ plane, where $\ell$ indexes their orbital angular momenta.  Likewise, let $\{\psi_\ell(\rhovec')\}$, for $\rhovec' = (x',y')$, be a set of orthonormal OAM modes on the circular receiver pupil $\mathcal{A}_L = \{\rhovec': |\rhovec'|\le D/2\}$ in the $z=L$ plane that are extracted by a mode converter in that pupil.  From the extended Huygens-Fresnel principle, we have that the complex field envelope, $\zeta_\ell(\rhovec')$, of the field produced in the $z=L$ plane by transmission of $\Psi_\ell(\rhovec)$ from $\mathcal{A}_0$ is
\begin{equation}
\zeta_\ell(\rhovec') = \int_{\mathcal{A}_0}\!d\rhovec\,\Psi_\ell(\rhovec)h_L(\rhovec',\rhovec),
\end{equation}
where $h_L(\rhovec',\rhovec)$ is the atmospheric Green's function.  The unnormalized average cross-talk between received OAM modes $\ell$ and $\ell'$ in the $\mathcal{A}_L$ pupil is therefore
\begin{equation}
C_{\ell,\ell'} \equiv \left\langle\left|\int_{\mathcal{A}_L}\!d\rhovec'\,\psi_{\ell'}^*(\rhovec')\zeta_\ell(\rhovec')\right|^2\right\rangle,
\label{crosstalk}
\end{equation}
where angle brackets denote averaging over the turbulence ensemble.  
It follows that $C_{\ell,\ell'}$ is completely characterized by the mutual coherence function of the atmospheric Green's function, viz., 
\begin{eqnarray}
C_{\ell,\ell'} &=& \int_{\mathcal{A}_L}\!d\rhovec_1'\int_{\mathcal{A}_L}\!d\rhovec_2'\int_{\mathcal{A}_0}\!d\rhovec_1 \int_{\mathcal{A}_0}\!d\rhovec_2\,\psi_{\ell'}(\rhovec_1')\psi_{\ell'}^*(\rhovec_2')\Psi_\ell^*(\rhovec_1)\Psi_\ell(\rhovec_2) 
\nonumber \\[.05in]
&\times& \langle h_L^*(\rhovec_1',\rhovec_1)h_L(\rhovec_2',\rhovec_2)\rangle.
\end{eqnarray}
For Kolmogorov-spectrum turbulence, we have that \cite{Shapiro,Yura,Ishimaru}
\begin{equation}
\langle h_L^*(\rhovec_1',\rhovec_1)h_L(\rhovec_2',\rhovec_2)\rangle = 
\frac{e^{-ik(|\rhovec_1'-\rhovec_1|^2 - |\rhovec_2'-\rhovec_2|^2)/2L}}{(\lambda L)^2}\,e^{-D(\rhovec_1'-\rhovec_2',\rhovec_1-\rhovec_2)/2},
\label{MCF}
\end{equation}
where $k = 2\pi/\lambda$ is the wave number, the fraction on the right is due to vacuum propagation, and 
\begin{eqnarray}
\lefteqn{D(\rhovec_1'-\rhovec_2',\rhovec_1-\rhovec_2) \equiv }\nonumber \\[.05in]
&& \hspace*{.5in}2.91 k^2\int_0^L\!dz\,C_n^2(z)|(\rhovec_1'-\rhovec_2')z/L + (\rhovec_1-\rhovec_2)(1-z/L)|^{5/3},
\label{structure}
\end{eqnarray}
is due to turbulence, whose strength profile along the path is $C_n^2(z)$.  The initial derivation of this mutual coherence function employed the Rytov approximation \cite{Shapiro,Yura}, hence $D(\rhovec_1'-\rhovec_2',\rhovec_1-\rhovec_2)$ was termed the two-source, spherical-wave, wave structure function, and the validity of (\ref{MCF}) and (\ref{structure}) was limited to the weak-perturbation regime before the onset of saturated scintillation.  Later \cite{Ishimaru}, it was shown that (\ref{MCF}) and (\ref{structure}) could be obtained from the small-angle approximation to the linear transport equation, making them valid well into saturated scintillation.  

Several key points are worth noting here.  First, non-uniform turbulence distributions lead to there being several coherence lengths of potential interest, including: (a)  the transmitter-pupil coherence length
\begin{equation}
\rho_0 \equiv \left(2.91k^2 \int_0^L\!dz\,C_n^2(z)(1-z/L)^{5/3}\right)^{-3/5},
\end{equation}
which quantifies the beam spread incurred in propagation from $z=0$ to $z=L$, because transmission of a complex field envelope $E_0(\rhovec)$ from $\mathcal{A}_0$ yields a complex field envelope $E_L(\rhovec')$ in $\mathcal{A}_L$ whose average irradiance is
\begin{eqnarray}
\lefteqn{\langle |E_L(\rhovec')|^2\rangle = }\nonumber \\[.05in]
&& \int_{\mathcal{A}_0}\!d\rhovec_1\int_{\mathcal{A}_0}\!d\rhovec_2\,E_0^*(\rhovec_1)E_0(\rhovec_2) 
\frac{e^{-ik(|\rhovec'-\rhovec_1|^2 - |\rhovec'-\rhovec_2|^2)/2L}}{(\lambda L)^2}\,e^{-(|\rhovec_1-\rhovec_2|/\rho_0)^{5/3}/2};
\end{eqnarray}
and (b), the receiver-pupil coherence length
\begin{equation}
\rho_0' \equiv \left(2.91k^2 \int_0^L\!dz\,C_n^2(z)(z/L)^{5/3}\right)^{-3/5},
\end{equation}
which quantifies the angle-of-arrival spread for a point-source transmission $E_0(\rhovec) = \delta(\rhovec)$ from $\mathcal{A}_0$, because a diffraction-limited, focal-length $f>0$ lens in $\mathcal{A}_L$ yields an average image-plane irradiance 
\begin{equation}
\langle |E_{L'}(\rhovec)|^2\rangle = \int_{\mathcal{A}_L}\!d\rhovec_1'\int_{\mathcal{A}_L}\!d\rhovec_2'\,\frac{e^{ik\rhovec\cdot(\rhovec_1'-\rhovec_2')/L'}}{(\lambda L)^2(\lambda L')^2}\,e^{-(|\rhovec_1'-\rhovec_2'|/\rho_0')^{5/3}/2},
\end{equation}
where $1/L' = 1/L - 1/f$. For a uniform distribution of turbulence---$C_n^2(z)$ constant from $z=0$ to $z=L$---these two coherence lengths coincide, 
\begin{equation}
\rho_0 = \rho_0' = (1.09 k^2 C_n^2 L)^{-3/5}.
\end{equation}

In addition to $\rho_0$ and $\rho_0'$, there is one more coherence length we need to introduce.  Suppose that a plane wave is transmitted from the $z=0$ plane, i.e., $E_0(\rhovec) = E_0$ for all $\rhovec$ in that plane.  
The mutual coherence function of the resulting $z=L$ field, found from (\ref{MCF}) and (\ref{structure}) with $\mathcal{A}_0$ extended to cover the entire $z=0$ plane, obeys
\begin{eqnarray}
\lefteqn{\langle E_L^*(\rhovec_1)E_L(\rhovec_2')\rangle = } \nonumber \\[.05in]
&& \hspace*{.35in} \int\!d\rhovec_1\int\!d\rhovec_2\, |E_0|^2\frac{e^{-ik(|\rhovec_1'-\rhovec_1|^2 - |\rhovec_2'-\rhovec_2|^2)/2L}}{(\lambda L)^2}\,e^{-D(\rhovec_1'-\rhovec_2',\rhovec_1-\rhovec_2)/2} = \nonumber \\[.05in]
&& \hspace*{.5in}|E_0|^2 e^{-D(\rhovec_1'-\rhovec_2',\rhovec_1'-\rhovec_2')/2},
\end{eqnarray}
for Kolmogorov-spectrum turbulence, where the second equality follows from integrating in sum and difference coordinates, $\rhovec_+ = (\rhovec_1+\rhovec_2)/2$ and $\rhovec_- = \rhovec_1-\rhovec_2$.  
Note that 
\begin{equation}
D(\rhovec_1'-\rhovec_2',\rhovec_1'-\rhovec_2') = \left(2.91 k^2 \int_0^L\!dz\,C_n^2(z)\right) |\rhovec_1'-\rhovec_2'|^{5/3},
\end{equation}
is the plane-wave structure function of the Rytov theory, whose coherence length is 
\begin{equation}
\rho_P = \left(2.91 k^2 \int_0^L\!dz\,C_n^2(z)\right)^{-3/5}
\end{equation}
in general, and 
\begin{equation}
\rho_P = (2.91 k^2 C_n^2L)^{-3/5}
\end{equation}
for uniformly-distributed turbulence.

\section{The Rodenburg {\em et al}~Simulator}
Rodenburg \em et al\/\rm\ \cite{Rodenburg} performed a laboratory experiment scaled to simulate propagation of 785-nm-wavelength OAM modes over a 1-km-long path of uniformly distributed, Kolmogorov-spectrum turbulence with $C_n^2 = 1.8 \times 10^{-14}\,{\rm m}^{-2/3}$.  In what follows, however, we shall stick with the unscaled path geometry, rather than the scaled version Rodenburg \em et al\/\rm\ used in their experiments.  

In \cite{Rodenburg}, the position and strength of the two phase screens were chosen to match the following propagation parameters for the propagation path specified above---the Fried parameter version of the plane-wave coherence length, $r_0 = (6.88)^{3/5}\rho_P$; the plane-wave log-amplitude variance, $\sigma^2_\chi = 0.31 k^{7/6} C_n^2 L^{11/6}$; the normalized variance, $\sigma^2_P$, of the power collected by the 18.2-cm-diameter receiver pupil from a plane-wave transmission; and the density of branch points in that receiver pupil, $\rho_{BP}$---see \cite{Rodenburg} for the details.  Rodenburg \em et al\/\rm\ give $r_0$ values and locations for the two phase screens they claim will simulate propagation through the thick turbulent path described above:  $r_{01} = 3.926\,$cm, $r_{02} = 3.503$\,cm, $z_1 = 171.7\,$m, and $z_2 = 1.538\,$m.   To connect with the extended Huygens-Fresnel principle theory laid out in section~2, we note that Rodenburg \em et al\/\rm's phase screens correspond to the impulsive $C_n^2(z)$ distribution
\begin{equation}
C_n^2(z) = N_{n1}^2\delta(z-z_1) + N_{n2}^2\delta(z-z_2),
\end{equation}
where
\begin{equation}
N_{nm}^2= 6.88/2.91 k^{2} r_{0m}^{5/3} = \left\{\begin{array}{ll}
8.14\times 10^{-12}\,{\rm m}^{-1/3}, &\mbox{for $m=1$}\\[.05in]
9.84\times 10^{-12}\,{\rm m}^{-1/3}, &\mbox{for $m=2$}.\end{array}\right.
\end{equation}

At this point it is easy to see the limitation of the Rodenburg \em et al\/\rm\ simulation.  Their two-screen $C_n^2(z)$ distribution yields
\begin{equation}
\rho_0 = \left[2.91 k^2 \left(N_{n1}^2(1-z_1/L)^{5/3} + N_{n2}^2(1-z_2L)^{5/3}\right)\right]^{-3/5} = 8.3\,{\rm mm},
\end{equation} 
and
\begin{equation}
\rho_0' = \left[2.91 k^2 \left(N_{n1}^2(z_1/L)^{5/3} + N_{n2}^2(z_2L)^{5/3}\right)\right]^{-3/5} = 7.19\,{\rm cm},
\end{equation} 
for the transmit and receive pupil coherence lengths,
whereas the uniformly-distributed turbulence they are trying to simulate would have
\begin{equation}
\rho_0 = \rho_0' = (1.09 k^2 C_n^2 L)^{-3/5} = 1.38\,{\rm cm}.
\end{equation}
Given the above coherence-length discrepancies, one cannot expect the Rodenburg \em et al\/\rm\ simulator to yield accurate results for $C_{\ell,\ell'}$ for all choices of the OAM modes $\Psi_\ell(\rhovec)$.  The question then becomes when could it provide an accurate cross-talk assessment?  Because the Rodenburg \em et al\/\rm\ simulator matched propagation parameters for a plane-wave source, it is reasonable to suggest collimated-beam OAM modes as the natural candidates for accurate $C_{\ell,\ell'}$ determination via that simulator.  Indeed, although \cite{Rodenburg} does not say so, the cross-talk results reported therein were obtained with precisely such modes \cite{RodenburgPrivate}, i.e., 
\begin{equation}
\Psi_\ell(\rhovec) = \sqrt{\frac{4}{\pi D^2}}\,e^{i\ell\phi}, \mbox{ for $|\rhovec|\le D/2$,}
\label{collimated}
\end{equation}
where $\phi$ is the azimuthal angle of $\rhovec$.

Because the $D\rightarrow\infty$ limit of (\ref{collimated}) is an OAM-modulated plane wave, we can expect that the $C_{\ell,\ell'}$ results from \cite{Rodenburg} should be accurate when $D$ is large enough that turbulence-induced beam spread can be ignored.  For a simple and optimisitic initial assessment of whether beam spread is insignificant in the \cite{Rodenburg} scenario, we will replace its 18.2-cm-diameter circular pupils with square pupils having 18.2\,cm sides and calculate
\begin{equation}
\mathcal{F}_0 \equiv \int_{-D/2}^{D/2}\!dx'\int_{-D/2}^{D/2}\!dy'\,\langle|\zeta_0(\rhovec')|^2\rangle
\end{equation}
for 
\begin{equation}
\Psi_0(\rhovec) = \frac{1}{D}, \mbox{ for $|x|,|y|\le D/2$}
\end{equation}
when $\rho_0 = 3.8\,$mm (the $z=0$ plane coherence length for the Rodenburg \em et al\/\rm\ simulator), and compare that result with the corresponding vacuum-propagation ($\rho_0 = \infty$) result.  After some algebra we get
\begin{eqnarray}
\mathcal{F}_0 &=& \int_{-1}^{1}\!dv_x\int_{-1}^1\!dv_y\, \frac{\sin[\pi \sqrt{D_f}\,v_x (1-|v_x|)]}{\pi v_x}\,
\frac{\sin[\pi \sqrt{D_f}\,v_y (1-|v_y|)]}{\pi v_y} \nonumber \\[.05in]
&\times&
\frac{\sin[\pi \sqrt{D_f}\,v_x]}{\pi \sqrt{D_f}\,v_x}\,\frac{\sin[\pi \sqrt{D_f}\,v_y]}{\pi \sqrt{D_f}\,v_y}\,
 e^{-(\sqrt{v_x^2+v_y^2}\,D/\rho_0)^{5/3}/2},
\label{power}
\end{eqnarray}
where $D_f = D^4/(\lambda L)^2$ is the Fresnel-number product of the transmitter-receiver geometry.  
Equation~(\ref{power}) yields $F_0 = 0.929$ for vacuum propagation and $F_0 = 0.719$ for the turbulent case.  Inasmuch as Rodenburg \em et al\/\rm's circular pupils inscribe our square pupils, and 
\begin{equation}
\Psi_\ell(\rhovec) = \frac{e^{i\ell \phi}}{D}, \mbox{ for $|x|,|y|\le D/2$}
\end{equation}
with $\ell \neq 0$ has higher spatial-frequency content than does $\Psi_0(\rhovec)$, we believe that the results from \cite{Rodenburg} are on the edge of providing an accurate cross-talk assessment.  A more definitive statement about the validity of \cite{Rodenburg} would require full comparison between its experimental results and numerical evaluation of (\ref{crosstalk}), using (\ref{MCF}) and (\ref{structure}) with the parameter values for the horizontal-path scenario \cite{Rodenburg} chose to simulate.

\section*{Conclusions}
We have shown that the two-screen turbulence simulator from \cite{Rodenburg} does not properly represent the Green's-function mutual coherence for a uniform distribution of Kolmogorov-spectrum turbulence.  As a result, the average cross-talk predictions from \cite{Rodenburg} for reception without adaptive optics---predictions that can be directly compared with those obtained from the extended Huygens-Fresnel principle---may only be valid for that paper's collimated-beam OAM modes when turbulence-induced beam spread is insignificant.   Moreover, if the results for cross-talk without adaptive optics are suspect, then those obtained for cross-talk mitigation with adaptive optics must also be regarded with skepticism.

\end{document}